\documentstyle[aps,12pt]{revtex}

\begin{document}
\author{O.B.Zaslavskii}
\address{Department of Physics, Svobody Sq.4, Kharkov State University, Kharkov\\
310077, Ukraine,\\
E-mail: olegzasl@aptm.kharkov.ua}
\title{Entropy of quantum fields for nonextreme black holes in the extreme limit}
\maketitle

\begin{abstract}
Nonextreme black hole in a cavity within the framework of the canonical or
grand canonical ensemble can approach the extreme limit with a finite
temperature measured on a boundary located at a finite proper distance from
the horizon. In spite of this finite temperature, it is shown that the
one-loop contribution $S_{q\text{ }}$of quantum fields to the thermodynamic
entropy due to equilibrium Hawking radiation vanishes in the limit under
consideration. The same is true for the finite temperature version of the
Bertotti-Robinson spacetime into which a classical Reissner-Nordstr\"{o}m
black hole turns in the extreme limit. The result $S_{q}=0$ is attributed to
the nature of a horizon for the Bertotti-Robinson spacetime.
\end{abstract}

\draft


The present paper is devoted to the description of thermodynamic properties
of black holes near the extreme limit in the framework of the Euclidean path
integral approach. There is good reason to believe that such properties
depend strongly on the topological sector to which the black hole belongs.
Therefore, truly extreme black holes and nonextreme ones near the extreme
state may have little in common. Thermodynamics in the first case was
considered in \cite{hawking} where it was argued that the zero-loop entropy $%
S$ for extreme black holes equals zero$,$ the horizon being situated
infinitely far away from any point outside it. The second situation was
discussed in \cite{zaslprl}, \cite{zasl97} for the black hole in a finite
size cavity in the framework of the grand canonical ensemble. Then the
entropy $S$ is equal to its standard Bekenstein-Hawking value $A/4$ ($A$ is
the surface area of the horizon). The aim of the present paper is to extend
the previous result indicated above and examine the possible role of quantum
corrections $S_{q}$ to the entropy for the limiting case considered in \cite
{zaslprl}. For this case the limit is taken in such a way that the
temperature measured on the boundary is finite and the proper distance
between the horizon and any other point inside a cavity stays finite. I show
that properties of the black hole in such an extreme limit are rather
unexpected: although the temperature $T\neq 0$, $S_{q}=0.$ In other words,
the entropy of a dressed black hole comes entirely from the spacetime
geometry ($S=A/4$) as it does for a bare hole.

The problem considered below has, apart from a general interest to
properties of near-extreme black holes, one more motivation connected with
the status of the third law of thermodynamics (Nernst theorem) in black hole
thermodynamics. (This issue has been recently touched upon in \cite{wald}).
This theorem asserts that the entropy should go to zero (or some universal
constant) when temperature approaches zero. The results of either the
previous paper \cite{zaslprl} or the present one show that one should be
very careful in the formulation of the third law in the context of black
hole thermodynamics due to the crucial difference between the Hawking
temperature $T_{H}$ and temperature $T$ which determines properties of the
gravitational ensemble and is imposed on the boundary of the cavity inside
which the hole is situated \cite{york86}. In the case under discussion the
zero-loop entropy $S\neq 0$, $S_{q}=0$; $T\neq 0,T_{H}=0.$

Throughout the paper we mean by $S_{q\text{ }}$the one-loop contribution of
a hot gas of quantum radiation. This quantity is finite, being the linear
functional of an renormalized stress-energy tensor $T_{\mu }^{\nu }$ in the
Hartle-Hawking state \cite{zasl93}, \cite{zasl96}. We do not calculate here
quantum corrections to the entropy of the black hole itself and do not
consider here in detail the mechanism of renormalization: it is assumed
right from the very beginning that such renormalization is already
performed, so we deal with the finite $S_{q}$ and $T_{\mu }^{\nu }$ in the
Hartle-Hawking state. Thus, the total entropy $S=A/4G+S_{q}$ where $G$ is
the renormalized gravitational constant. In particular, it means that
information about the structure of divergent corrections to the
statistical-mechanical entropy \cite{frolov} of the black hole itself near
the extreme state \cite{solod95}, \cite{demers},\cite{cognola} is
essentially insufficient for determining $S_{q}.$

The proof of the statement that $S_{q}=0$ in the limit under discussion is
based on recovering $S_{q}$ from $T_{\mu }^{\nu }$. Under some physically
reasonable restrictions on the behavior of the metric near the horizon we
derive the formula for $S_{q}$ in terms of $T_{\mu }^{\nu }$ for a
spherically-symmetrical black hole spacetime which probably can be useful
for further applications as well. After this formula is obtained for an
arbitrary relation between $r_{+}$ and $r_{B}$ the limiting transition to $%
r_{+}=r_{B}$ corresponding to the state under consideration is made.

This limiting transition is the characteristic feature of the configuration
under discussion. The points with $r>r_{+}$ of the original spacetime (say,
of the Reissner-Nordstr\"{o}m (RN) black hole) lie at infinite proper
distance from $r_{+},$ as usual. However, since the relevant manifold is
obtained in the limiting case under discussion by suitable expanding the
vicinity of the horizon into the region with a finite Euclidean four-volume,
all points of this manifold pick up the value $r=r_{+}$ for which the metric
coefficient $g_{rr}\rightarrow \infty .$ As a result, the proper distance $%
l=\int_{r_{+}}^{r}dr\sqrt{g_{rr}}$ is finite for the configuration at hand
(see \cite{zaslprl}, \cite{zasl97} for details).

We suggest also another approach to the problem. It is shown in \cite{zasl97}
that the limiting geometry of a RN black hole near this state coincides with
the Bertotti-Robinson (BR) spacetime with a non-zero surface gravity.
Starting from the BR spacetime right from the very beginning, we show that $%
S_{q}=0$ in agreement with the result obtained by the limiting transition
from the RN metric. Bearing in mind physical relevance of BR spacetimes as
useful tool for discussing many aspects of the Hawking effect \cite{lapedes}%
, this result can be of interest on its own.

Let us consider the Euclidean RN metric 
\begin{equation}
ds^{2}=fd\tau ^{2}+f^{-1}dr^{2}+r^{2}d\omega ^{2}  \label{metric}
\end{equation}

$d\omega ^{2}=d\phi ^{2}\sin ^{2}\theta +d\theta ^{2},$ $%
f=(1-r_{+}/r)(1-e^{2}/r_{+}r).$

Here the Euclidean time $0\leq \tau \leq \beta _{0}=\beta _{H\text{ }}$
where the inverse Hawking temperature 
\begin{equation}
\beta _{H}=4\pi r_{+}(1-e^{2}/r_{+}^{2})^{-1}  \label{Hawking T}
\end{equation}

At finite temperature the one-loop action $I_{q}$ of quantum fields has the
standard form \cite{dowker} 
\begin{equation}
I_{q}=-\int d^{4}x\sqrt{g}T_{0}^{0}-S_{q}  \label{action}
\end{equation}

which for the metric \ref{metric} reduces to 
\begin{equation}
I_{q}=\beta _{H}M_{q}-S_{q},  \label{action sph}
\end{equation}

$M_{q}=-4\pi \int\limits_{r_{+}}^{r_{B}}drr^{2}T_{0}^{0}.$ It is assumed
that the black hole is enclosed in the cavity of a radius $r_{B}.$
Properties of quantum fields in thermal equilibrium between a black hole and
its Hawking radiation are described by the renormalized stress-energy tensor
in the Hartle-Hawking state. In this state all diagonal components of $%
T_{\mu }^{\nu }$ are regular at the horizon (off-diagonal components for our
metric are equal to zero).

Now we will recover $S_{q}$ from $T_{\mu }^{\nu }.$ Let us consider the
variation of the metric between two equilibrium states with different $r_{+}$
but the same $e.$ One cannot use the standard formula 
\begin{equation}
\delta I_{q}=\frac{1}{2}\int d^{4}x\sqrt{g}T_{\mu \nu }\delta g^{\mu \nu }
\label{del}
\end{equation}
directly, since when $r_{+}$ varies the limits of integration either in $%
\tau $ or in $r$ change themselves along with the metric. Therefore, it is
convenient to introduce new variables which take their values between fixed
limits. Let $\tau =\tilde{\tau}\beta _{H}/2\pi $, $%
y=(r-r_{+})/(r_{B}-r_{+}). $ Then $0\leq \tilde{\tau}\leq 2\pi $, $0\leq
y\leq 1.$ Now we may use the above equation immediately. In so doing, 
\begin{equation}
\delta g^{yy}=(\frac{\delta f}{f}+\frac{2\delta r_{+}}{r_{B}-r_{+}})g^{yy},%
\text{ }\delta g^{00}=-(\frac{\delta f}{f}+\frac{2\delta \beta _{H}}{\beta
_{H}})g^{00},\text{ }\delta g^{aa}=\frac{2(y-1)}{r}g^{aa}\delta r_{+}
\label{var}
\end{equation}
where $a=\theta ,\phi .$

As $f=f(r,r_{+})=f(r(r_{+},y),r_{+})$ we have (the dependence on $e$ is
omitted for shortness) to take into account the full dependence of $f$ on $%
r_{+}$ in calculating $\delta f$. Then, substituting \ref{var} into \ref{del}
we obtain 
\begin{equation}
\delta I_{q}/4\pi =-\delta \beta
_{H}\int_{r_{+}}^{r_{B}}drr^{2}T_{0}^{0}+\delta r_{+}\beta
_{H}/2\int_{r_{+}}^{r_{B}}drr^{2}h(T_{r}^{r}-T_{0}^{0})+\gamma \beta
_{H}\delta r_{+}  \label{delta I}
\end{equation}
where $h=f^{-1}\frac{\partial f}{\partial r_{+}},$ $\gamma
=\int_{r_{+}}^{r_{B}}drr^{2}\{T_{y}^{y}y^{\prime }+(1-y)[f^{\prime
}(T_{y}^{y}-T_{0}^{0})/2f-(T_{\theta }^{\theta }+T_{\phi }^{\phi })/r]\},$
prime denotes derivative with respect to $r$.

Now take into account that $T_{y}^{y}=T_{r}^{r}$ and make use the
conservation law according to which

\begin{equation}
r^{-2}(T_{r}^{r}r^{2})^{\prime }=(T_{0}^{0}-T_{r\,})f^{\prime
}/2f+r^{-1}(T_{\theta }^{\theta }+T_{\phi }^{\phi })  \label{conserv}
\end{equation}

Then $\gamma =\int_{r_{+}}^{r_{B}}dr[T_{r}^{r}r^{2}(y-1)]^{\prime
}=T_{r}^{r}(r_{+})r_{+}^{2}$ as $y=1$ at $r=r_{B}$ and $y=0$ at $r=r_{+}.$
Thus, eventually we have 
\begin{equation}
\delta I_{q}/4\pi =-\delta \beta
_{H}\int_{r_{+}}^{r_{B}}drr^{2}T_{0}^{0}+\delta r_{+}\beta
_{H}T_{r}^{r}(r_{+})r_{+}^{2}+\delta r_{+}\beta
_{H}/2\int_{r_{+}}^{r_{B}}drr^{2}h(T_{r}^{r}-T_{0}^{0})  \label{delta final}
\end{equation}

For the Schwarzschild metric ($e=0$) one can use the scale properties of it
and after integration in \ref{delta final} reproduce the result of \cite
{zasl93} and \cite{zasl96}. It gives $S_{q}$ in terms of one-dimensional
integrals from $T_{\mu }^{\nu }.$ Although for the RN metric the expression
for $S_{q}$ is more complicated, it is rather convenient for analysis.

Comparing \ref{action sph} and \ref{delta final} one can get

\begin{equation}
\frac{\partial }{\partial r_{+}}S_{q}=-4\pi \beta _{H}g(r_{+},r_{B}),\
g=g_{1\,+}g_{2},\text{ }g_{1}=\int_{r_{+}}^{r_{B}}drr^{2}\frac{\partial }{%
\partial r_{+}}T_{0}^{0},\text{ }g_{2}=1/2%
\int_{r_{+}}^{r_{B}}drr^{2}h(T_{r}^{r}-T_{0}^{0})  \label{entropy der}
\end{equation}

Integrating \ref{entropy der}, we can write down 
\begin{equation}
S_{q}=4\pi \int_{r_{+}}^{r_{B}}d\tilde{r}_{+}\beta _{H}(\tilde{r}_{+})g(%
\tilde{r}_{+},r_{B})  \label{entropy}
\end{equation}

The constant of integration is chosen to make sure $S_{q}\rightarrow 0$ when 
$r_{+}\rightarrow r_{B}$ for a generic nonextreme hole: when there is no
room for radiation its entropy must be zero.

The expression \ref{entropy} for the entropy is a linear functional of $%
T_{\mu \nu }.$ In the Hartle-Hawking state $T_{\mu }^{\nu }=(T_{\mu }^{\nu
})_{th}+(T_{\mu }^{\nu })_{B}$ where ($T_{\mu }^{\nu })_{B}$ is the
stress-energy tensor in the Boulware state, ($T_{\mu }^{\nu })_{th}$ $%
=\alpha T^{4}diag(-1,1/3,1/3,1/3)$ is that of thermal radiation \cite{tensor}%
. Therefore, the entropy under discussion $S_{q}=S_{th}+S_{B}$ where $S_{th}$
is the entropy of thermal radiation (statistical-mechanical entropy of the
gas of massless quanta) and $S_{B\text{ }}$is the renormalization constant.
Both $S_{th}$ and $S_{B}$ are divergent but their sum is finite due to the
finiteness of the $T_{\mu }^{\nu }$ in the Hartle-Hawking state. Such
splitting explains the renormalization procedure for the thermal atmosphere
of a black hole developed in \cite{zurek} and shows that the entropy under
discussion harmonizes with the statistical-mechanical entropy after such a
renormalization (cf. discussion in \cite{frolov} where, however, the
relevance of splitting properties of $T_{\mu }^{\nu }$ in the Hartle-Hawking
state were not pointed out).

The limit we are operating with is just $r_{+}=r_{B}.$ However, the state
the black hole approaches is now {\it extreme} $(m=e=r_{+}).$ For such a
state the Euclidean four-volume is finite \cite{zaslprl}, \cite{zasl97} and
whether $S_{q}=0$ or $S_{q}\neq 0$ is not obvious in advance. Indeed, we
deal with the competition of two factors: the integration region shrinks $%
(r_{+}\rightarrow r_{B})$ but $\beta _{H}(r_{+})\rightarrow \infty $ (cf.
the calculation of the proper distance between the horizon and boundary in 
\cite{zaslprl}). It is the evaluation of the expression \ref{entropy} that
we now turn to.

As far as $g_{1}$ is concerned, it is essential that the geometry of the
spacetime is regular everywhere. More precisely, it takes the form of the
Bertotti-Robinson metric \cite{zasl97} (see below). Correspondingly, $T_{\nu
}^{\mu }$ approach their BR values. Therefore, the integrand in $g_{1}$
remains finite and $\lim_{r_{+}\rightarrow r_{B}}g_{1}=0.$ Consider the
behavior of $g_{2}.$ For the RN metric the quantity $%
h=-r^{-1}(1-r_{+}/r)^{-1}(1-e^{2}/r_{+}r)^{-1}(1-e^{2}/r_{+}^{2})\propto
(1-r_{+}/r)^{-1}$when $r\rightarrow r_{+}.$ The regularity condition at the
horizon demands $T_{0}^{0}=T_{r}^{r}$ at $r=r_{+}.$ Therefore, $%
g_{2}\rightarrow 0$ along with $g_{1\text{ }}.$ Thus, $\lim_{r_{+}%
\rightarrow r_{B}}g(r_{+,}r_{B})=0.$

To show that $S_{q}=0$ it is sufficient to prove that $\lim_{r_{+}%
\rightarrow r_{B}}\int_{r_{+}}^{r_{B}}d\tilde{r}_{+}\beta _{H}(\tilde{r}%
_{+})\equiv C<\infty .\,$Direct calculation with \ref{Hawking T} taken into
account shows that 
\begin{equation}
C=2\pi r_{B}^{2}\lim \ln \left[ (r_{B}-e)/(r_{+}-e)\right]  \label{c}
\end{equation}

Up to this point we only used the limiting relation $r_{+}=m=e=r_{B}$
between parameters but did not consider in which way they approach this
limit. Now we specify it by demand that our limit correspond to the finite
local temperature on the boundary. Then, as was shown in \cite{zaslprl}, $%
e=pxr_{+},$ $x=r_{+}/r_{B}$ where the parameter $p=1-\varepsilon ,$ $%
x=1-\varepsilon \alpha ,$ $\varepsilon \rightarrow 0$ but $\alpha $ remains
the finite nonzero quantity. Either $p$ or $\alpha $ can be expressed in
terms of boundary data but their concrete values are irrelevant for our
purposes. The only conclusion to be drawn from these properties is that
either the numerator or the denominator in \ref{c} inside the logarithm have
the same order in $\varepsilon $, so their ratio is a finite number. Thus, $%
C<\infty $ and $S_{q}=0.$ It is worth noting that although $M_{q}\rightarrow
0$ according to \ref{action sph}, the product $M_{q}\beta _{H}\propto
(r_{B}-r_{+})/(r_{+}-e)$ remains finite for the same reason, so $I_{q}\neq
0. $

As follows from the method of derivation, the result $S_{q}=0$ as the black
hole approaches the extreme state $(T_{0}\rightarrow 0)$ in the non-extreme
topological sector remains valid for the more general type of metrics: 
\begin{equation}
ds^{2}=fd\tau ^{2}+V^{-1}dr^{2}+r^{2}d\omega ^{2}  \label{more general}
\end{equation}
What is needed is only the character of asymptotic behavior: $f\propto
A(r-r_{+})(r-r_{-}),$ $V\propto B(r-r_{+})(r-r_{-})$ near the horizon ($A$
and $B$ are constants, $r_{-}$ is the radius of an inner horizon), the limit
under discussion corresponding to $r_{+}=r_{-}.$

As a matter of fact, the above proof relied on general properties of
spacetime and regularity of $T_{\mu }^{\nu }$ near the horizon but did not
use explicitly the concrete form of the limiting geometry. Its character was
not obvious in advance since the coordinate $r$ becomes singular and $%
f\rightarrow 0.\,$It was shown in \cite{zasl97} that the metric \ref{metric}
turns in this limit into 
\begin{equation}
ds^{2}=r_{+}^{2}(d\tau ^{2}\sinh ^{2}x+dx^{2}+d\omega ^{2})  \label{bert}
\end{equation}
which is the version of the BR spacetime \cite{bertotti}, \cite{robinson}
with the non-zero surface gravity. Here $0\leq \tau \leq 2\pi .$ Bearing in
mind the physical significance of the issue of quantum corrections to the
black hole geometry near the extreme state, below we present another proof
of the property $S_{q}=0$ starting from \ref{bert} right from the very
beginning. As it relies on \ref{bert} directly independently of where it
originates from, such a proof can be of interest for studying thermal
properties of the BR spacetime itself.

The quantum stress-energy tensor of massless scalar fields regular at the
horizon reads \cite{and4}, \cite{kofman} 
\begin{equation}
T_{\mu }^{\nu }=r_{+}^{-4}A\delta _{\mu }^{\nu }  \label{tensor}
\end{equation}
where $A$ is a pure number whose value is irrelevant for our purposes. Let
us consider the finite portion of \ref{bert} with the boundary at $x=x_{B}.$
Then the action is the function of two independent variables $x_{B}$ and $%
r_{+}$. Now we can recover the action from the stress-energy tensor in the
manner similar to that described above. As the metric \ref{bert} differs
from \ref{metric}, I sketch below the main points of calculations.

As explained above, one cannot use directly eq.\ref{del} since the
integration region itself changes under variation of $x_{B}.$ Let us
introduce the variable $y$ according to $x=yx_{B},$ $0\leq y\leq 1.$ Then eq.%
$\ $\ref{del} is applicable. Let $x_{B}$ be varied while keeping $r_{+}$
fixed. Then after simple transformations one gets 
\begin{equation}
(\frac{\partial I}{\partial x_{B}})_{r_{+}}=-8\pi ^{2}\sinh x_{B}  \label{16}
\end{equation}
Integrating this equation we obtain 
\begin{equation}
I=-8\pi ^{2}(\cosh x_{B}-\xi )  \label{17}
\end{equation}
Here the quantity $\xi $ is in fact constant not depending on $r_{+}$ as it
follows from dimension grounds. One should put $\xi =1$ that the action obey
the boundary condition $I(x_{B}=0)=0\,$: when there is no room for radiation
its action must be zero. Then the action \ref{17} can be rewritten in the
form 
\begin{equation}
I_{q}=-r_{+}^{-4}\int d^{4}x\sqrt{g}A  \label{action bert}
\end{equation}
where $g$ is the determinant of the metric \ref{bert}, $0\leq x\leq x_{B},$ $%
0\leq \tau \leq 2\pi .$ One can easily check that the action \ref{action
bert} reproduces correctly conformal anomaly of the tensor \ref{tensor}. For
this purpose it is necessary to take into account that this action depends
on $r_{+}$ either via the metric or via the factor $r_{+}^{-4}$ as a
parameter. Then the variation of the metric leads to the conformal anomaly
term which is compensated by the variation of the $r_{+}^{-4}$ factor to
give $\frac{\partial I}{\partial r_{+}}=0$ in agreement with eq.\ref{17}.

Comparing eq.\ref{action bert} with the general form \ref{action} and taking
into account \ref{tensor}, we conclude that $S_{q}=0$ in agreement with the
above calculations for RN black holes. It is worth noting that although back
reaction of quantum fields will certainly change the geometry \ref{bert} it
will be the effect of the next order in the Planck constant for $S_{q}$, so
at least in the one-loop approximation (quantum fields propagating on a
classical background) $S_{q}=0.$

The above result $S_{q}=0$ admits qualitative physical explanation. Let us
pass from \ref{bert} to its Lorentzian version by substitution $\tau =it.$
Then introducing new variables according to 
\begin{equation}
t_{1}=r_{+}e^{t}\coth x,\text{ }\rho =r_{+}e^{t}(\sinh x)^{-1}
\label{new variables}
\end{equation}
we obtain the metric in the form 
\begin{equation}
ds^{2}=\rho ^{-2}r_{+}^{2}(-dt_{1}^{2}+d\rho ^{2}+\rho ^{2}d\omega ^{2})
\label{new br}
\end{equation}

The metric \ref{bert} is equivalent only to the part $t_{1}>\rho $ of the
metric \ref{new br} that resembles the relation between the Rindler space
(analog of \ref{bert}) and the Minkowski one (analog of \ref{new br}). The
event horizon of an accelerated observer in the flat spacetime represents
the pure kinematic effect in the sense that by suitable coordinate
transformation the metric turns into explicitly flat form and does not have
an event horizon for an inertial observer at infinity in contrast to black
hole spacetimes. For this reason, in spite of effects of interaction between
an accelerated detector and Rindler quanta \cite{birrel} the renormalized
tensor $T_{\mu }^{\nu }=0$ in the Hartle-Hawking state which corresponds in
this case to the Minkowski vacuum of an inertial observer \cite{ginz}, so $%
S_{q}=0$ for the thermodynamic entropy in the Rindler space. (To avoid
confusion, I recall that I discuss only finite renormalized quantities which
contribute to thermodynamics and do not consider in detail the mechanism of
compensation between divergent zero-point vacuum oscillations in curved
background and thermal excitations.)

It goes without saying that relationship between \ref{bert} and \ref{new br}
is more complicated. In particular, the BR spacetime has three non-commuting
Killing vectors, the metric \ref{new br} is geodesically incomplete itself 
\cite{lapedes}, etc. However, one can think that qualitatively the reason
for $S_{q}=0$ is the same: the horizon in the BR spacetime \ref{bert} is not
a true black hole horizon and does not reveal itself in $S_{q}$ which, being
coordinate independent quantity, can be calculated in the reference frame 
\ref{new br} where the horizon at $\rho =\infty $ is ''too weak'' ($T_{H}=0$%
) to gain entropy of quantum fields.

One can look at the result $S_{q}$ from the more general viewpoint. Let we
have a static spacetime $M_{1}.$ Perform transformation from the original
reference frame ($x^{i},t)$ in which the metric does not depend on time to
the new one $(\tilde{x}^{i},\tilde{t}),$ both coordinate systems being
nondegenerate. Let the metric in the new frame also do not depend on time,
the coordinates $(\tilde{x}^{i},\tilde{t})$ covering only the part $M_{2}$
of $M_{1}.$ Let the spacetime $M_{1}$ has the horizon for observers
following orbits of the timelike Killing vector such that the associated
Hawking temperature $T_{H}=0$ or does not have a horizon at all whereas
there is a horizon in $M_{2}$ with the corresponding Hawking temperature $%
\tilde{T}_{H}\neq 0$. The relationship between two sections of the BR
spacetime corresponds to the first case while the relationship between the
Minkowski and Rindler spacetimes corresponds to the second one. It is
plausible that the property $S_{q}=0$ reflects the kinematic nature of a
horizon in $M_{2}$ in contrast to the black hole case and follows directly
from the general assumptions sketched above. It is of interest to elucidate
whether this statement can be proven rigorously without the reference to an
explicit form of the metric.

{\it Note added .} In the recent paper \cite{mann} R.B.Mann and
S.N.Solodukhin considered quantum corrections to the entropy of a black hole
itself and the mechanism of renormalization in the extreme limit in question
and found that such corrections due to massless scalar fields have the
universal behavior. In fact, the result $S_{q}=0$ for the entropy of Hawking
radiation derived in our paper can also be considered as a manifestation of
universality inherent to the extreme limit.

\end{document}